\documentclass[proceedings]{JHEP} % 10pt is ignored!

\usepackage{epsfig}			% please use epsfig.
\def\be{\begin{equation}}
\def\ee{\end{equation}}
\def\bea{\begin{eqnarray}}
\def\eea{\end{eqnarray}}

\def\mxth{\mathsurround=0pt }
\def\xversim#1#2{\lower2.pt\vbox{\baselineskip0pt \lineskip-.5pt
x  \ialign{$\mxth#1\hfil##\hfil$\crcr#2\crcr\sim\crcr}}}

\renewcommand{\a}{\alpha}

\newcommand{\pa}{\partial}

\newcommand{\q}{\theta}

\newcommand{\Ka}{{K\"ahler}}

\newcommand{\ft}[2]{{\textstyle\frac{#1}{#2}}}

\newcommand{\eqn}[1]{(\ref{#1})}
\newbox\mybox
\newcommand\fverb{\setbox\mybox=\hbox\bgroup\verb}
\newcommand\fverbdo{\egroup\medskip\noindent\fbox{\unhbox\mybox}\ }
\newcommand\fverbit{\egroup\item[\fbox{\unhbox\mybox}]}

\font\beeg=cmr17 scaled 1600		% Stylish initials
\newcommand\init[1]{\setbox\mybox=\hbox{{\beeg #1}~}%
		   \noindent\global\hangindent=\wd\mybox\global\hangafter-2%
		   \sc\smash{\llap {\lower 13.2pt \box\mybox}}}
\title{INSTANTONS AND QUATERNIONS}

\author{Stefan Vandoren\\
	YITP, SUNY at Stony Brook, Stony Brook, NY 11794-3840, USA\\
	E-mail: \email{vandoren@insti.physics.sunysb.edu}}

\conference{Nonperturbative Quantum Effects, 2000}

\abstract{We relate the moduli space of Yang-Mills instantons to
quaternionic manifolds. For instanton number one, the Wolf spaces play 
an important role. We apply these ideas to instanton calculations in 
${\cal N}=4$ SYM theory.}

\begin{document} 

\maketitle %%%%%%%%%% THIS IS IGNORED %%%%%%%%%%%

% Here we start...

Instantons in Yang-Mills theories are defined by the solutions to the
self-duality equation in four-dimensional Euclidean space,
\be
F_{\mu\nu}={}^*F_{\mu\nu}=
\ft12 \epsilon_{\mu\nu\rho\sigma}F_{\rho\sigma}\ ,\label{inst}
\ee
and are characterized by an integer number, the topological charge,
\be
k=-\frac{1}{16\pi^2}\int\,{\rm d}^4x\, {\rm Tr} \,F_{\mu\nu}{}^*F_{\mu\nu}\ .
\ee
For $k=1$, one can explicitly solve these 
equations \cite{BPST,H}, and for gauge group $SU(2)$, the instanton 
gauge field one-form $A$ and its field strength $F$ can be elegantly 
written in terms of quaternions
$x=x_\mu \sigma_\mu$, with $\sigma_\mu=(\vec \tau,i)$, \cite{At},
\bea
A&=&i\,{\rm Im}\,\frac{({\bar x}-{\bar x}_0){\rm d}x}{\rho^2+|x-x_0|^2}\nonumber\\
F&=&\frac{2\rho^2{\rm d}{\bar x}\wedge {\rm d}x}{(\rho^2+|x-x_0|^2)^2}\ ,
\eea
where the norm of quaternions is defined as $|x^2|=\ft12{\rm tr}(x{\bar x})
=x_\mu x_\mu$. This solution has five collective coordinates, four 
positions $x_0^\mu$ and
a size $\rho$, but one can act with rigid $SU(2)$ gauge rotations with 
angles ${\vec \theta}$ to generate new inequivalent solutions. 
So in this example, there are eight collective coordinates. 

The purpose of this letter is to elaborate on the geometry of the moduli 
space of collective coordinates, and to apply this knowledge to 
instanton calculations in ${\cal N}=4$ supersymmetric Yang-Mills theories.

For $k=1$ and gauge group $SU(2)$, the moduli space is 
simply ${\bf R}^8$ (in fact, it is $ {\bf R}^4\times {\bf
R}^4/Z_2$ , which  has an orbifold singularity corresponding to
zero size instantons), with metric
\be
\label{SU2metric}
{\rm d}s^2={\rm d}x_0^\mu{\rm d}x_0^\mu+({\rm d}\rho)^2+\rho^2{\rm d}S^3\ .
\ee
This metric can be computed by evaluating the inner product of zero modes, 
which are the derivatives of the instanton field with respect to the 
collective coordinates, as was demonstrated in 
\cite{H,Bernard}~\footnote{For a recent review on instantons, and how to 
determine the zero modes and metric, see e.g. \cite{BVV}.}.
We have dropped certain normalization
factors, which are irrelevant for the present discussion. 
The last term in \eqn{SU2metric} comes from the gauge orientation 
zero modes, whose corresponding
collective coordinates ${\vec \theta}$ parametrize $SU(2)\equiv S^3$.

For $k=1$ and gauge group $SU(N)$, one can construct all instanton
 solutions by embedding
the $SU(2)$ instanton inside $SU(N)$, and acting with rigid gauge 
transformations on the chosen embedding,
\bea
\label{SUN-inst}
A^{SU(N)}_\mu&=&U(\phi)\pmatrix{0 & 0 \cr 0 & A_\mu^{SU(2)}}U^\dagger(\phi)
\nonumber\\
&&\nonumber\\
 U&\in & \frac{SU(N)}{SU(N-2)\times U(1)}\ .
\eea
Notice that we have divided by the stability group of the instanton embedding.
The $SU(N-2)$ factor in \eqn{SUN-inst} only acts on the upper diagonal
block and commutes with the $SU(2)$ embedding. Furthermore, there is an
extra singlet, corresponding to the $U(1)$ factor in the stability group,
commuting with $SU(2)$ ({\it e.g.}, for $SU(3)$ with the standard 
Gell-Mann matrices, the first three determine the $SU(2)$ embedding, whereas 
the eighth generator commutes with this embedding and is the $U(1)$ singlet
mentioned above).

The dimension of this coset space is $4N-5$, and we denote the 
coordinate-angles by $\phi=\{\vec \theta,q_i^M\}, i=1,...,N-2$, $M=1,...,4$
 so counting leads to $4N$ collective coordinates in total, in agreement with 
standard index theorems.

The metric on the moduli space of $k=1$ $SU(N)$ instantons is \cite{Bernard} 
\be
\label{SUNmetric}
{\rm d}s^2={\rm d}x_0^\mu{\rm d}x_0^\mu+({\rm d}\rho)^2+
\rho^2{\rm d}S^3+\rho^2{\rm d}X_{N-2}\ ,
\ee
where ${\rm d}X_{N-2}$ stands for the metric element on the
$4(N-2)$-dimensional space
\be
X_{N-2}=\frac{SU(N)}{SU(N-2)\times SU(2)\times U(1)}\ .
\ee
Actually this metric is the one obtained by working infinitesimally
in the angles $\phi$, for which it is known that the ${\vec \theta}$ 
zero modes are orthogonal to the $q_i^M$ zero modes, see for instance
\cite{BVV}. For reasons to be explained below, one expects that the 
full metric in \eqn{SUNmetric} is of the form where $S^3$ is fibered 
non-trivially over $X_{N-2}$.

The space  $X_{N-2}$ is known as one of the  Wolf spaces \cite{Wolf} 
and is an example of a quaternionic manifold, which we will 
discuss later. The angles $q_i^M$ can then be interpreted as $N-2$
quaternions, which coordinatize the quaternionic manifold. 
Ignoring the center of 
mass coordinates $x_0^\mu$, the metric has the form of a cone, with radial
parameter $\rho$.

For $k=1$ and gauge group $SO(N)$, there are $4N-8$ collective coordinates.
The one-instanton solution is constructed by choosing 
an $SU(2)$ inside an $SO(4)=SU(2)\times SU(2)$ embedding 
in $SO(N)$ \cite{BCGH}. The stability group of this instanton is $SO(N-4)
\times SU(2)$, and performing a similar counting as for the $SU(N)$ case,
one indeed obtains $4N-8$ collective coordinates.
The metric on the moduli space is now
\be
\label{SONmetric}
{\rm d}s^2={\rm d}x_0^\mu{\rm d}x_0^\mu+({\rm d}\rho)^2+
\rho^2{\rm d}S^3+\rho^2{\rm d}Y_{N-4}\ ,
\ee
where ${\rm d}Y_{N-4}$ stands for the metric element on the 
$4(N-4)$-dimensional Wolf space
\be
Y_{N-4}=\frac{SO(N)}{SO(N-4)\times SO(4)}\ .
\ee

Finally we analyze the gauge group $Sp(N)$. Here we can simply choose the lower
diagonal $SU(2)=Sp(1)$ embedding inside $Sp(N)$. The stability group
of this embedding is now $Sp(N-1)$, and the moduli space metric is
\be
\label{SpNmetric}
{\rm d}s^2={\rm d}x_0^\mu{\rm d}x_0^\mu+({\rm d}\rho)^2+
\rho^2{\rm d}S^3+\rho^2{\rm d}HP_{N-1}\ ,
\ee
with corresponding $4(N-1)$-dimensional quaternionic projective space
\be
HP_{N-1}=\frac{Sp(N)}{Sp(N-1)\times Sp(1)}\ .
\ee
Counting leads to $4(N+1)$ collective coordinates, which agrees again with
index theorems. One could do a similar analysis for the exceptional
gauge groups, and obtain a corresponding quaternionic manifold. 
It is known that for any semisimple Lie algebra, there 
is a corresponding quaternionic Wolf space \cite{Wolf}.
In general a quaternionic manifold (in the mathematics literature called
quaternionic-\Ka\, although they are not of the K\"ahlerian type)
is a $4n$-dimensional Riemannian space with holonomy group contained
in $Sp(1)\times Sp(n)$. These spaces are always Einstein. The 
classification of the compact homogeneous quaternionic manifolds
is precisely given by the Wolf spaces \cite{Wolf,Aleks}.
Hence the $k=1$ instanton moduli spaces are characterized by the Wolf series.
Quaternionic
spaces also appear in the coupling of four-dimensional ${\cal N}=2$ 
hypermultiplets to supergravity \cite{BagWit}. 
We will discuss its relation to instantons later.

For arbitrary instanton number $k$, one can not construct the instanton
 solution
explicitly. However, using the  ADHM formalism (which extensively uses 
a quaternionic notation), it is possible 
to write down the solutions in an implicit way \cite{ADHM}. 
The moduli space, denoted by ${\cal M}_k(G)$, has dimension $4kN, 4k(N-2)$ and
$4k(N+1)$ for gauge groups $SU(N), SO(N)$ and $Sp(N)$ respectively.
It is known to be a hyper-\Ka\ manifold of a special kind \cite{Mac,Swann},
hence also called {\it special hyper-\Ka\ manifolds} \cite{DWKV}, because 
it has the following properties:

\begin{itemize}

\item The metric on the moduli space, $g$, allows for a 
hyper-\Ka\ potential, $\chi$, which satisfies, in local coordinates,
\be
g_{AB}=D_A\pa_B \chi\ .
\ee
Maciocia \cite{Mac} has shown that for Yang-Mills instantons,
\be
\chi=\frac{1}{16\pi^2}\int\,{\rm d}^4x\, |x|^2\, {\rm Tr}\,
F_{\mu\nu}F_{\mu\nu}\ ,
\ee
and for $k=1$, one can explicitly compute this potential, $\chi=2\rho^2+
|x_0|^2$.

\item The derivative $\chi_A=\partial_A \chi$ is a homothetic Killing vector,
and there is an $SU(2)$ isometry, generated by 
\be 
{\vec k}^A={\vec J}^A{}_B \chi^B\ ,
\ee
where ${\vec J}$ are the three complex structures of the hyper-\Ka\ 
manifold. This $SU(2)$ isometry rotates the complex structures.
Four-dimensional hyper-\Ka\ spaces with such an $SU(2)$ isometry
are classified. They are either flat space, Taub-Nut, or the Atiyah-Hitchin
manifold. Only flat space allows for a homothetic Killing vector.

\item One can factor out the center of mass coordinates of the $k$-instanton
solution, such that 
\be
{\cal M}_k={\bf R}^4\times {\cal M}^0_k\ .
\ee
Here ${\cal M}^0_k$ is the reduced instanton moduli space. It is again
a special hyper-\Ka\ manifold \cite{BoMa}.
Spaces with a homothety can always be described as a
cone. One can choose $\chi$ as one of the coordinates, and denote the 
remaining coordinates by $x^{\hat A}$. The line element
can then be written in the form \cite{GibbonsRych},  
\be
{\rm d}s^2 = {{\rm d}\chi^2\over 2 \chi} +  \chi\,
h_{\hat A\hat B}(x) \,{\rm d}x^{\hat A}\,{\rm d}x^{\hat B} \,, \label{conemetr}
\ee
where the $x^{\hat A}$ are the coordinates associated with the
hypersurface $\chi={\rm constant}$. It is known that this hypersurface
 must be a
3-Sasakian space and the hyper-\Ka\ space is therefore a cone
over the 3-Sasakian metric. For a recent review on these spaces, see
\cite{sasaki}.
  
For $k=1$ one can see the cone structure in the reduced moduli space metrics
\eqn{SUNmetric},\eqn{SONmetric} and \eqn{SpNmetric}.
Indeed, the hyper-\Ka\ potential is, up to normalization, 
just the size of the instanton, $\chi=\rho^2$. 
At $\rho=0$, there is a conical singularity, corresponding
to zero size instantons. 

\item It turns out that 3-Sasakian spaces are $Sp(1)$ fibrations 
 over quaternionic \Ka\ manifolds ${\bf Q}$ \cite{Swann}. Hence 
one can write
\be
{\cal M}_k^0={\bf R}^+ \times [ Sp(1) \times _f {\bf Q}_k]\ ,
\ee
where ${\bf R}^+$ is parametrized by the cone variable $\chi$, and 
the subscript $f$ denotes the non-trivial fibration.
This structure is indeed present in the above discussed examples.
For $k=1$, ${\bf Q}_1$ are precisely the Wolf spaces, which are the only compact
homogeneous quaternionic manifolds. For higher instanton number, ${\bf Q}_k$
will in general not be homogeneous anymore, and it will contain 
singularities. It would be interesting to 
study its properties. 

\end{itemize}

In an apparently different context, special hyper-\Ka\ manifolds
also appear as target spaces for four-dimensional (but also in five 
or six dimensions~\footnote{For two-dimensional sigma models on group
manifolds, with eight supercharges, one also recovers the 
Wolf spaces \cite{SSTV}.})
superconformal hypermultiplets, as was
discussed in detail in \cite{DWKV,dubna,dWRV}, see also \cite{Galicki}. 
There, the implications of rigid ${\cal N}=2$ superconformal invariance
on sigma models were analyzed, and the constraints on 
the hyper-\Ka\ manifold are exactly the ones discussed above.
The corresponding quaternionic space then appears if one
would couple the hypermultiplets to supergravity via the superconformal 
tensor calculus \cite{STC}.
This raises the question if the instanton moduli space naturally
appears as the target space of some sigma model theory. This is indeed
the case, as was shown in \cite{Witt,Dougl} and also in \cite{DHKMV}.
There, instantons are described in terms of Dp-D(p+4) brane 
configurations, and the instanton moduli space coincides
with the vacuum moduli space of the corresponding worldvolume theory
of the Dp-D(p+4) system.

To make this correspondence more clear, we should consider instantons
in ${\cal N}=4$ SYM theory at the conformal point ({\it i.e.}, not 
spontaneously broken), and for simplicity we
choose gauge group $SU(N)$ and instanton number $k=1$. 
On top of the $4N$ bosonic zero modes,
we have $8N$ fermionic zero modes, each one corresponding to a solution
of the Dirac equation. The Grassmannian collective coordinates (GCC)
are denoted by
\be
 \xi^M_\a,{\bar \eta}^{M\dot \a}, \mu^M_i, {\bar \mu}^{Mi}\ ,
\ee
see for instance \cite{DKMV}, where $M=1,...,4$ is the $SU(4)_R$ symmetry
index, $\a,\dot \a=1,2$ are two-component spinor indices, and $i=1,...,N-2$
is related to the color index. The first sixteen of these GCC, $\xi^M$ and
${\bar \eta}^M$ are related to supersymmetry and superconformal 
supersymmetry, {\it i.e.}, their zero modes can be obtained by acting with 
the super(conformal-) symmetry transformations on the bosonic instanton
configuration. Only half of the symmetries are broken, so for ${\cal N}=4$
we get $8+8=16$ fermionic zero modes. 

In instanton calculations, one is interested in the instanton action.
This is obtained by plugging in the instanton configuration (which in 
${\cal N}=4$ SYM is only an approximate solution to the field equations)
into the SYM action and integrating over four-dimensional space. 
The resulting action only depends on the collective coordinates, and for 
$k=1$, $SU(N)$ SYM, it is \cite{DKMV},
\be
S_{\rm inst}=-\frac{\pi^2}{4g^2\rho^2}
\epsilon_{MNPQ}\,\mu^M_i{\bar \mu}^{Ni}\,\mu^P_j
{\bar \mu}^{Qj}\ ,\label{mu-four}
\ee
where $g^2$ is the YM coupling constant.

The remaining of this paper is to understand this result in terms
of superconformal hypermultiplet sigma models, and to point out the relation
with the quaternionic geometry described above.
All the collective coordinates are organized in hypermultiplets, and
we have $N$ of them. Two hypermultiplets are special, they comprise the
eight scalars $x_0^\mu,\rho, {\vec \theta}$ and fermions $\xi^M,
{\bar \eta}^M$. This ``universal'' sector appears in 
every SYM theory, and is independent of the gauge group. 
The fermions of the universal sector will not appear in the
instanton action, since they are protected by super(conformal-)symmetry.
The other hypermultiplets contain the quaternionic gauge orientation 
zero modes $\q_i^M$ and fermions $\mu^M_i,{\bar \mu}^{Mi}$. 
The latter correspond
to fermionic zero modes which can not be obtained by acting with 
fermionic symmetry transformations on the bosonic instanton configuration.
Hence one might expect them in the instanton action, as \eqn{mu-four}
indeed confirms.

The instanton moduli space as a Higgs branch was first discovered in the 
D5-D9 system in type I string theory \cite{Witt}. The corresponding 
world-volume theory on the Higgs branch is a sigma model in six dimensions
with eight rigid supercharges. The general action for such a 
system contains 
kinetic terms for the scalars and fermions, and a four-fermi term 
coupled to the Riemann tensor of the $4n$-dimensional 
hyper-\Ka\ target space \cite{ST},
\be
\label{four-fermi}
S_{4-{\rm fermi}}\propto R_{abcd}\,
\epsilon_{MNPQ}\,\psi^{Ma}\psi^{Nb}\psi^{Pc}\psi^{Qd} \ .
\ee
Here $\psi^{M}$ is a six-dimensional $SU(4)\approx SO(6)$ symplectic 
Majorana spinor, $\epsilon_{MNPQ}$ is the totally 
antisymmetric $SU(4)$ invariant tensor, $a=1,...,2n$ is the $Sp(n)$ index, and 
$R_{abcd}$ is the totally symmetric $Sp(n)$ curvature of the 
hyper-\Ka\ target space, which has holonomy group contained 
in $Sp(n)$, see \cite{BagWit,JBBS} for more details. The instanton action
in \eqn{mu-four} can be compared with the system in \eqn{four-fermi}
by dimensional reduction to zero dimensions.
 In terms of string theory, we
are then describing the D(-1)-D3 system, which is well known to represent
a four-dimensional SYM instanton \cite{Dougl}. After dimensional 
reduction, all 
kinetic terms have disappeared and one is left with only the Riemann tensor as
in \eqn{four-fermi}. Applied to the $k=1$ $SU(N)$ instanton model, we 
have $n=N$, and the $8N$ fermions $\psi^{Ma}$ comprise $\xi^M_\a,
{\bar \eta}^{M\dot \a},\mu^M_i,{\bar \mu}^{Mi}$. Now, because the 
$\xi$ and ${\bar \eta}$ zero modes are protected by super(conformal-)symmetry,
they cannot appear in the instanton action. So the only surviving components
are the ones where the Riemann tensor has its indices in the subspace 
$Sp(N-2)\subset Sp(N)$. This is in accord with the general structure presented
above. First one factors out the centre of mass coordinates, such that
the reduced instanton moduli space has holonomy contained in $Sp(N-1)$.
As was shown in \cite{DWKV}, the $Sp(N-1)$ curvature of this special 
hyper-\Ka\ space has a quaternionic zero eigenvector. This corresponds
to the fact that the superconformal zero modes do not appear in the instanton 
action.
Secondly, this reduced moduli space is a cone over an $Sp(1)$ fibration
of a $4(N-2)$-dimensional quaternionic manifold, with holonomy group
contained in $Sp(1)Sp(N-2)$. Its coordinates are given by $q_i^M$
and we expect that it is precisely the $Sp(N-2)$ curvature which
couples to the four-fermi term containing the $\mu^{M}_i$'s.
For $k=1$, one can compue the Riemann tensor of the instanton moduli space
explicitly, it is essentially determined by the Riemann tensor of the 
corresponding Wolf
space. Since these spaces are symmetric, there is a basis of coordinates
in which all components of the curvature tensor (with two upper and two 
lower indices) are constant and determined by the structure constants of 
the coset space. This is consistent with \eqn{mu-four}, although the 
details of the computations remain to be worked out. We hope to come
back to this in the future.

To conclude, we believe that the relation
between instanton moduli spaces and quaternionic geometry is worth pursuing.

\acknowledgments

I would like to thank the organizers, in particular Bernard Julia,
for inviting me to the conference, and for the warm and pleasant atmosphere 
during that week. I also acknowledge discussions with
Tim Hollowood, V. V. Khoze and M. Trigiante.

%%%%%%%%%%%%%%%%%%%%%%%%%%%%%%%%%%%%%%%%%%%%%%%%%%%%%%%%%%%%%%
%%%%%%%%%%%%%%%%%%%%%%%%%%%%%%%%%%%%%%%%%%%%%%%%%%%%%%%%%%%%%% 
% ---- Bibliography ----
%

\end{document}